\documentclass[aps,prl,twocolumn,showpacs,superscriptaddress, floatfix]{revtex4-2}
\usepackage{amsmath, amssymb, graphicx, bm, subcaption, mathtools}

\usepackage[usenames,dvipsnames,svgnames,table]{xcolor}
\usepackage[colorlinks=true,linkcolor=RoyalBlue,citecolor=RoyalBlue, urlcolor=RoyalBlue]{hyperref}
\begin{document}

	\title{Deterministically switching the Bloch chirality of domain walls and skyrmions: \\ the Dzyaloshinskii astroid}
	
	\author{Michael D.  Kitcher}
	\email{mkitcher@andrew.cmu.edu}
	\affiliation{Department of Materials Science \& Engineering, Carnegie Mellon University, Pittsburgh, PA 15213, USA}
	
	\author{Larry Chen}
	\affiliation{Department of Materials Science \& Engineering, Carnegie Mellon University, Pittsburgh, PA 15213, USA}
	
	\author{Marc De Graef}
	\affiliation{Department of Materials Science \& Engineering, Carnegie Mellon University, Pittsburgh, PA 15213, USA}
	
	\author{Vincent Sokalski}
	\affiliation{Department of Materials Science \& Engineering, Carnegie Mellon University, Pittsburgh, PA 15213, USA}
	
	\begin{abstract}
		
		%%We present a mechanism for deterministic control of the Bloch chirality in magnetic skyrmions originating from the combination of an interfacial Dzyaloshinskii-Moriya interaction (DMI) and a perpendicular magnetic field. Although the conventional interfacial DMI is known to favor the formation of chiral N\'eel skyrmions, it does not break the energetic symmetry of the two possible Bloch chiralities in mixed Bloch--N\'eel skyrmions.  However, the energy barrier to switching between Bloch chiralities does depend on the sense of rotation, which can be controlled by the direction of a driving field. %The results of this analysis lead to the formation of a 
		%Our analysis culminates in a switching diagram for Dyaloshinskii domain walls---akin to the Stoner--Wohlfarth astroid---and reveals the existence of different Bloch chirality regimes.
		%%Introducing the concept of domain wall susceptibility, our analysis of steady-state Dzyaloshinskii DW dynamics culminates in a switching diagram---akin to the Stoner--Wohlfarth astroid---and reveals the existence of both chiral and achiral Bloch regimes. Furthermore, we discuss recent theory of vertical Bloch line--mediated Bloch chirality selection in the precessional regime and extend these arguments to vertical Bloch line evolution at lower driving fields. This work establishes that effective magnetic fields, which could have myriad origins, can be used to dynamically switch between chiral Bloch states as indicated by this new \textit{Dzyaloshinskii astroid}.
		
		We present a mechanism for deterministic control of the Bloch chirality in magnetic domain walls and skyrmions---originating from the interplay between an interfacial Dzyaloshinskii--Moriya interaction (DMI) and a perpendicular magnetic field. Although conventional interfacial DMI favors chiral N\'eel skyrmions, it does not break the energetic symmetry of the two Bloch chiralities in mixed Bloch--N\'eel skyrmions.  However, the energy barrier to switching between Bloch chiralities does depend on the sense of rotation, which is dictated by the direction of the driving field. Our analytical and micromagnetic analyses of steady-state Dzyaloshinskii domain wall dynamics culminates in a switching diagram akin to the Stoner--Wohlfarth astroid, revealing the existence of achiral, chiral, and precessional Bloch regimes; we further show that in view of established energetic models, skyrmions should exhibit the same fundamental behavior.  Finally, we discuss a recent theory of vertical Bloch line--mediated deterministic Bloch chirality in the precessional regime and extend these arguments to lower driving fields. This work establishes that the Bloch chirality of magnetic solitons can be switched on demand with applied fields, as indicated by the \textit{Dzyaloshinskii astroid}.

	\end{abstract}
	
	%CROSS-CHECK REFERENCES
	%ADD UPDATED FIGURES AND CAPTIONS
	
	\maketitle
	
	%Introduction
	%	Background on how chirality is controlled
	%	Absence of any work on controlling chirality

	The internal structure of magnetic skyrmions and domain walls (DWs) is central to the blooming field of chiral magnetism \cite{Back2020}. The origin of this preferred chirality is the Dzyaloshinskii-Moriya interaction (DMI) \cite{Dzyaloshinsky1958,Moriya1960}, which can favor chiral N\'eel DWs in conventional ferromagnet/heavy metal (FM/HM) thin films with structural inversion asymmetry \cite{Thiaville2012, Hrabec2014}, as well as chiral Bloch DWs in B20 magnets which possess bulk inversion asymmetry \cite{Muhlbauer2009, Nagaosa2013}. Surprisingly, there have been relatively few reported studies of switching between chiral states, which could %serve as the basis for
	enable new binary magnetic memory schemes. These efforts have focused on the direct control of the strength and sign of interfacial DMI in FM/HM films via an applied electric field, so as to reverse the N\'eel chirality of skyrmions \cite{Nawaoka2015, Srivastava2018, Schott2021}.  %Even in this case
	However, the desired switched states would remain only if the electric field is applied perpetually.
	
	In %the present 
	this work, we consider switching between the Bloch states of DWs and skyrmions in FM/HM films with perpendicular magnetic anisotropy (PMA) and a conventional interfacial DMI, where the latter preferentially stabilizes one N\'eel handedness over the other but creates no difference in energy between the two Bloch %configurations
	chiralities.  %The
	This switching process is driven by an effective out-of-plane magnetic field %,
	which reorients the internal magnetization of a DW or skyrmion to a steady-state configuration. The pinning-free dynamics of field-driven DWs were first formalized and investigated by Walker, Slonczewski, and Schryer in a series of seminal papers \cite{Dillon1963, slonczewski1972, Schryer1974}. %If the effective field is large enough,
	Without DMI, field-based switching was only expected beyond the critical Walker field where a DW undergoes Walker breakdown: the in-plane magnetization of the wall %reorientation between
	cycles continuously %perpetual cycling
	through Bloch states, concurrent with a reduction in the DW's velocity.  Here, we show that for films with interfacial DMI, a range of effective field strengths exist in the steady-state regime where the driving torque enables unidirectional switching of DWs from one Bloch chirality to another.  Importantly, the final chirality remains after the effective field is removed and the system returns to static equilibrium.

	%Approach
	%	Governing equation for DW energy
	%	Plot of DW energy noting the equivalent Bloch energies
	%	Note that the energy barrier depends on the sense of rotation
	
	%	Governing equations for steady-state configuration and restoring torque
	
	%Introduction of the Dzyaloshinskii astroid
	
	\begin{figure*}[!htbp]
		\includegraphics[width = 7 in]{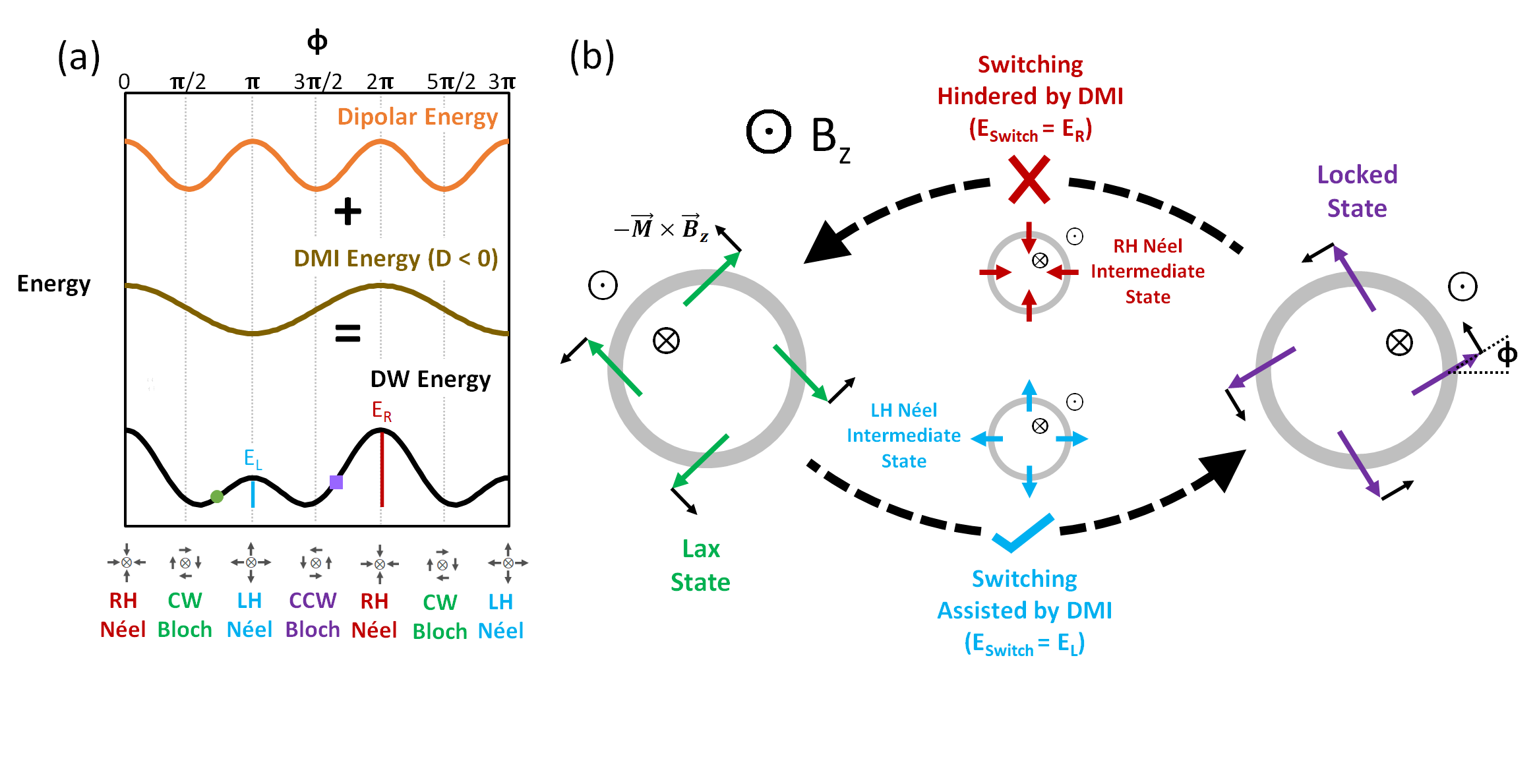}
		\caption{(a) %Representative
			Plots of the dipolar energy, DMI energy, and total energy versus $\phi$ for a Dzyaloshinskii DW %or skyrmion 
			where %$0 < \lvert{D}\rvert < D_c$
			$D < 0$. $E_L$ and $E_R$ are the energy barriers to switching between the two Bloch chiralities. For $B_z > 0$, the lability points of the lax and locked walls are labeled with round (green) and square (purple) markers, respectively. (b) Illustration of unidirectional switching between Dzyaloshinskii skyrmions for $0 < \lvert{D}\rvert < D_c$ and $B_{SW} < B < B_W$. For $D < 0$ and $B_z > 0$, DMI promotes the reorientation of a skyrmion with CW Bloch chirality (green moments; the laxed configuration) to one with a CCW Bloch chirality (purple moments; the locked configuration) via the preferred N\'eel chirality; a transition back to the lax %configuration 
			wall, however, is blocked. \small\label{Fig:StaticsAndDynamics}}
	\end{figure*}
	
	%\begin{figure}[!htbp]
	%\includegraphics[width = 0.4\textwidth]{DWEnergy.PNG}
	%\caption{(a) Schematic of mixed skyrmion with the defining coordinate system for the in-plane orientation of domain wall moments, $\phi$. (b) Plots of the dipolar energy, DMI energy, and the total energy versus $\phi$ for a Dzyaloshinskii domain wall where $D < 0$. $E_L$ and $E_R$ are the energy barriers to switching between the two Bloch wall variants. For $B_z > 0$, the lability points of the lax and locked walls are labeled with round (green) and square (purple) markers, respectively.\small\label{Fig:DWEnergy}}
	%\end{figure}
	
	%\begin{figure*}[!htbp]
	%\includegraphics[height = 4 in]{Dynamics.png}
	%\caption{Illustration of unidirectional switching between mixed skyrmions for $0 < \lvert{D}\rvert < D_c$ and $B_{SW} < B < B_W$. For $D < 0$ and $B_z > 0$, DMI promotes a dynamic reorientation of a skyrmion with clockwise Bloch chirality (green moments; the laxed wall configuration) to one with a counterclockwise Bloch chirality (purple moments; the locked wall configuration) via the preferred N\'eel chirality; a transition back to the lax wall, however, is blocked. \small\label{Fig:Dynamics}}
	%\end{figure*}
	
	The analyses in this work are built upon %proven analytical models of the domain wall
	the well-known q--$\phi$ collective coordinates model for magnetic domain walls. Here, the wall energy is given by the sum of the resting Bloch wall energy ($\sigma_0$), domain wall anisotropy energy, and DMI energy \cite{Thiaville2012, Pellegren2017}:
	\begin{subequations}
		\begin{align}
			\sigma &= \sigma_0 + 2{K_d}{\lambda_0}\cos{^2\phi} -{\pi}D\cos{\phi}\label{eq:sigma};\\
			\lim_{t_f \to 0} K_d &= \frac{\ln{[2]}{t_f}{\mu_0}{M_s}^2}{{2\pi\lambda_0}}\label{eq:Kd},
			%\lambda_0 &= \sqrt{A/K_{eff}}\label{eq:lambda0},
		\end{align}
	\end{subequations}
	where $D$ is the DMI constant, $K_d$ is the domain wall anisotropy, $t_f$ is the film's, $M_s$ is the saturation magnetization, $\lambda_0=\sqrt{A/K_{\text{eff}}}$ is the width of the DW, $A$ is the exchange stiffness, $K_{\text{eff}}$ is the effective PMA, $\mu_0$ is the vacuum permeability, and $\phi$ is the in-plane angle between the moment and normal of the DW. (See Figure \ref{Fig:StaticsAndDynamics}b.) As noted in Figure \ref{Fig:StaticsAndDynamics}a, a positive DMI constant stabilizes DW moments that point towards a $-M_z$ domain, i.e., a right-handed (RH) N\'eel DW. Conversely, $D < 0$ favors a left-handed (LH) N\'eel DW. The clockwise (CW) and counterclockwise (CCW) Bloch windings are likewise defined with respect to a $-M_z$ domain.
	
	It is well known that the DW takes on a fully N\'eel configuration when $\lvert{D}\rvert \geq D_c$ \cite{Thiaville2012, Cheng2019}, where
	\begin{align}
		D_c &= \frac{4\lambda_0K_d}{\pi};\nonumber\\ &= \frac{2\ln{[2]}}{\pi^2}{t_f}{\mu_0}{M_s}^2, t_f \to 0\label{eq:Dc};
	\end{align}
	
	In this work, we are focused on the regime where $\lvert{D}\rvert < D_c$ and the DW assumes a mixed Bloch--N\'eel configuration; the N\'eel component of the wall has a preferred chirality, whereas both winding directions are equally favored for its Bloch component. The corresponding existence of two energy minima is captured by \eqref{eq:sigma}, as illustrated in Figure \ref{Fig:StaticsAndDynamics}a. (For clarity, the $\phi$-independent contribution of the intrinsic Bloch energy is not included in the plots.) The in-plane orientation of a DW at static equilibrium (i.e. in its ground state), which is obtained when $\sigma_{\phi}$ = 0 and $\sigma_{\phi\phi}$ $>$ 0, %(subscripts of $\sigma$ denote partial derivatives)
	is then
	\begin{subequations}
		\begin{align}
			\cos{\phi_G}  &= m_{NG} = \frac{D}{D_c},\qquad\lvert{D}\rvert \leq D_c;\nonumber\\
			&= \text{sign}[D],\qquad\qquad\lvert{D}\rvert \geq D_c;\label{eq:nG}\\
			\sin{\phi_G} &= m_{BG} = {\pm}\sqrt{1 -(m_{NG})^2 },\label{eq:bG}
		\end{align}
	\end{subequations}
	where $m_{NG}$ and $m_{BG}$ are the respective N\'eel and Bloch components. As such, $m_{NG}$ also equals the relative strength of DMI with respect to the DW anisotropy energy. 
	
	We now turn our attention to the dynamic response of DWs subject to an effective perpendicular field.  Derived from the Landau--Lifshitz--Gilbert (LLG) equation, the Slonczewski equations of motion for the velocity and internal magnetization of DWs \cite{slonczewski1972, SanchezTejerina2017} are:
	\begin{subequations}
		\begin{align}
			\dot{\phi} &= {\frac{\gamma}{1 + \alpha^2}}(-\alpha\Omega + B_z)\label{eq:phiDot};\\
			\dot{q} &= {\frac{\gamma\lambda_0}{1 + \alpha^2}}(\Omega + \alpha{B_z})\label{eq:qDot};\\
			\Omega &= \frac{\sigma_{\phi}}{{2\lambda_0}M_s}\label{eq:restoringTorque},
		\end{align}
	\end{subequations}
	where $q$ is the position of the DW, $\gamma$ is the gyromagnetic ratio, $\alpha$ is the Gilbert damping constant, $B_z$ is the normalized driving torque, and $-\alpha\Omega$ represents the restoring torque that emerges as moments are driven away from static equilibrium by some $\Delta\phi$. 
	
	The steady-state configuration is therefore achieved when these two forces are balanced. From \eqref{eq:sigma}, the normalized restoring torque, $\sigma_\phi$, emerges as
	\begin{equation}
		\sigma_\phi = -4{K_d}{\lambda_0}\sin{\phi}\cos{\phi} + {\pi}D\sin{\phi}.\label{eq:sigmaPhi}
	\end{equation}
	At steady state, rearranging \eqref{eq:phiDot} yields
	\begin{equation}
		\sigma_{\phi} = \frac{2\lambda_0B_z{M_s}}{\alpha}%; \qquad\dot{\phi} = 0
		.\label{eq:restoringTorqueSteady}
	\end{equation}
	We can further describe the dynamic response of a DW to a driving field by introducing the concept of \textit{domain wall susceptibility} ($\chi_{dw}$): 
	\begin{equation}
		\chi_{dw} = \left(\frac{\partial B_z}{\partial \phi}\right)^{-1}.\label{eq:dwSusc}
	\end{equation}
	Combining \eqref{eq:restoringTorqueSteady} and \eqref{eq:dwSusc} at steady state, %the steady-state \textit{$\chi_{dw}$} is then
	\begin{equation}
		\chi_{dw} \propto {\sigma_{\phi\phi}}^{-1},% \qquad\dot{\phi} = 0, \label{eq:dwSuscFull}    
	\end{equation}
	where the proportionality constant is $\frac{2\lambda_0{M_s}}{\alpha}$ and  $\sigma_{\phi\phi}$ is the corresponding rate of change of the scaled restoring torque with respect to $\phi$. \newline
	
	%At small fields, in the absence of DMI, there are two solutions to this originating from the two degenerate static Bloch states (see figure xyz). 
	In the absence of DMI, there are two stable DW configurations in the steady-state regime with opposite Bloch chiralities, stemming from the corresponding degeneracy of the two static wall variants. (See the top plot of Figure \ref{Fig:StaticsAndDynamics}a.) At a critical $B_z$---conventionally termed the Walker field---both DWs experience the maximum restoring torque and $\lvert\Delta\phi\rvert$ possible at steady state, being stabilized at equivalent inflection points in the energy landscape (i.e., where $\chi_{dw} \to \infty$ and the moments of the DW are \textit{labile}, or most inclined to rotate). For $B_z$ values infinitesimally higher than this \textit{lability field}, the restoring torques cannot completely match the driving torque and both walls enter a precessional regime where their  %internal
	in-plane magnetizations rotate continuously. This behavior is accompanied by a sudden drop in the time-averaged velocity of each DW, as revealed in studies of domain expansion \cite{Metaxas2007}.
	
	Turning to the case of moderate interfacial DMI ($0 < \lvert{D}\rvert < D_c$, $0 < \lvert{m_{NG}}\rvert < 1$; see Figure \ref{Fig:StaticsAndDynamics}a), a third possibility emerges for moderate fields: a single steady-state solution \cite{SanchezTejerina2016, SanchezTejerina2017}. The evolution of the helicity of these Dzyaloshinskii DWs can be understood in the context of the aforementioned framework. Although both Bloch chiralities remain degenerate at static equilibrium, %the energy barrier associated with the transition from CW to CCW is significantly different from the CCW to CW transition if we fix the sense of rotation during the transition. Similarly, the restoring torque that balances the driving field torque in steady-state is larger for the CW to CCW transition.
	the energetic equivalence of the two N\'eel chiralities is broken. Beyond a shift in the energy minima towards the favored N\'eel chirality at static equilibrium, the presence of DMI also manifests as an asymmetry in the energy maxima which correspond to N\'eel configurations and act as barriers to switching between the two ground states.  On account of DMI, the two walls will therefore experience different net restoring torques when driven away from static equilibrium by the same effective field, leading to disparate variations in $\chi_{dw}$ during the process. As depicted in Figure \ref{Fig:StaticsAndDynamics}b, the wall that is driven towards the DMI-preferred N\'eel configuration---which we term the \textit{lax wall}---experiences an assistive torque from DMI, whereas the driving torque on its counterpart---the \textit{locked wall}---is opposed by the same effective torque generated by DMI. Consequently, the lability field is chirality dependent: the lax wall has the lower critical $B_z$ and larger $\Delta\phi$ at its lability point; Figure \ref{Fig:StaticsAndDynamics}a illustrates the case of $B_z < 0$. For driving fields in between these two limiting values, this wall will therefore adopt the steady-state configuration---and the Bloch chirality---of the locked wall. We can then characterize the lability field of the lax wall as the DW switching field ($B_{SW}$), while that of the locked wall remains the Walker field ($B_W$). From this treatment, %we can see that it is possible for 
	it is evident that a single Bloch chirality
	%to be 
	could be favored in magnetic systems with PMA and interfacial DMI when an effective $B_z$ is present.
	
	Within the framework of the q--$\phi$ DW energy model and the first Slonczewski equation (captured in \eqref{eq:sigma} and \eqref{eq:phiDot}, respectively), we then determine how $B_{SW}$ and $B_W$ depends on $D$. Using the steady-state restoring torque in \eqref{eq:restoringTorqueSteady}, the lability condition $\sigma_{\phi\phi}$ = 0, and normalized D ($m_{NG}$), we obtain the Néel and Bloch  wall moments ($m_{NL}$ and $m_{BL}$, respectively) at labile orientations ($\phi_L$):
	\begin{subequations}
		\begin{align}
			m_{NL} = \cos{\phi_L} &= \frac{m_{NG}\pm \sqrt{(m_{NG})^2 + 8}}{4}\label{eq:NL};\\
			m_{BL} = \sin{\phi_L} &= \mp\sqrt{(m_{NL}-m_{NG})m_{NL}}\nonumber\\
			&= \mp\sqrt{1-(m_{NL})^2}\label{eq:BL},
			%m_{BL} &= \sin{\phi} = \mp\sqrt{1 -(m_{NL})^2}\label{eq:BL}.
		\end{align}
	\end{subequations}
	where \eqref{eq:NL} corresponds to the expression for $\cos{\phi_W}$ in \cite{SanchezTejerina2016}. Here and henceforth, upper and lower signs correspond to expressions for lax and locked walls, respectively, for $D > 0$ (and vice versa for $D < 0$). %We note that \eqref{eq:NL} is equal to the expression for $\cos{\phi_W}$ in \cite{SanchezTejerina2016}, first found for the locked wall case, reduces to the expression for $\cos{\phi|_W}}$ found in \cite{Thiaville2012}.
	Combining these equations with \eqref{eq:sigmaPhi}, we obtain the critical ratio of driving field to Gilbert damping needed to drive each wall to a labile state: 
	\begin{align}
		\frac{B_z}{\alpha} &= \text{sgn}[B_z]\frac{2{K_d}}{M_s}(m_{NG} -m_{NL})m_{BL}\nonumber\\
		&=\text{sgn}[B_z]{\frac{K_d}{8M_s}}\sqrt{{{\pm{m_{NG}}+\sqrt{(m_{NG})^2 + 8}}}}\nonumber\\
		&\quad\times\sqrt{\left({\mp3m_{NG}+\sqrt{(m_{NG})^2 + 8}}\right)^3}.
		\label{eq:CBA}
	\end{align}
	
	\begin{figure*}[!htbp]
		\includegraphics[width = 7 in]{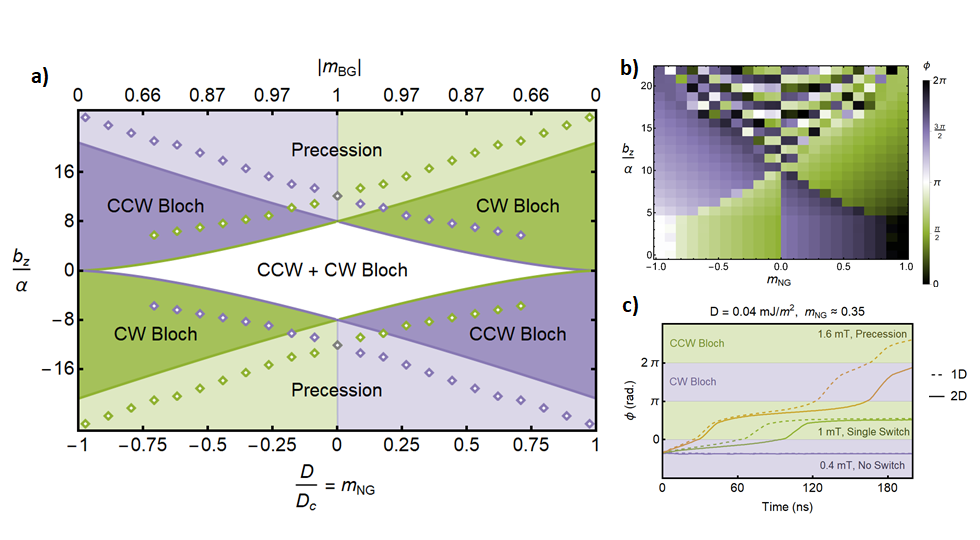}
		\caption{(a) The Dzyaloshinskii astroid derived from the q--$\phi$ model depicting the achiral Bloch (white), chiral Bloch (dark green and purple), and precessional (light green or purple) regimes for Dzyaloshinskii DWs and skyrmions as a function of reduced DMI and reduced driving field ($m_{NG}$ and $\frac{b_z}{\alpha}$, respectively). Corresponding values of $m_{BG}$, the normalized Bloch component of magnetization, are also indicated. The corresponding lability lines from 1D micromagnetic simulations (points) have been superimposed on the analytical result. (b) 1D Micromagnetic simulations of the field-driven evolution of Dzyaloshinskii DWs clockwise Bloch chirality with varying DMI strengths. (c) 2D and 1D micromagnetic simulations of counterclockwise Dzyaloshinskii DWs with $D > 0$ showing the field-driven time evolution of $\phi$ for representative cases of the achiral, chiral and precessional regimes. \small\label{Fig:Astroid}}
	\end{figure*}
	
	The results of this analysis are presented in the form of a DW phase diagram that indicates the critical field required to %reverse the Bloch chirality.
	stabilize a single Bloch chirality for a given DMI strength.  By normalizing \eqref{eq:CBA} by $\frac{K_d}{8M_s}$ to obtain the reduced quantity $\frac{b_z}{\alpha}$ and plotting the result as a function of $m_{NG}$ (which is also the reduced DMI), we obtain pairs of lability lines that delineate the various chirality regimes. For the case of $D = 0$, %we note that 
	the system directly transitions from the achiral Bloch regime to the precessional regime.  As %$\lvert D\rvert$
	the strength of DMI increases, a window emerges where there is a single steady-state solution (i.e., only one sense of switching is possible) before transitioning to the precessional regime.  The %full diagram 
	chiral region bears a striking resemblance to the Stoner--Wohlfarth astroid for single-domain switching first proposed by Slonczewski \cite[Ch.~3]{Hubert2008}. As the astroid in Figure \ref{Fig:Astroid}a characterizes the field-induced switching behavior of Dzyaloshinskii DWs, we propose to term it the  \textit{Dzyaloshinskii astroid}. In contrast to the referenced model of single-domain switching, the lability lines retain their physical relevance beyond the switching astroid, becoming the bounds of the precessional regime. As observed in \cite{Thiaville2012}, the Bloch component, $m_{BG}$, of the DW moments at static equilibrium is dominant for DMI values as high as $D = 0.7D_c$, which also lends the Dzyaloshinskii astroid to verification via Lorentz transmission electron microscopy \cite{Pandey2021}.  
	%Discussion
	%	Proposal to experimentally evaluate this phenomenon with perpendicular field
	%	Possible origin of preferred Bloch chirality in other work
	%		Refer to advanced materials paper to explain why this could work (creep, etc.)
	
	To support these analytical predictions and offer some approximate timescale over which the switching process would occur, we carried out micromagnetic simulations of the field-induced dynamics of 1D and 2D domain walls. Using MuMax3, we numerically solved the Landau--Lifshitz--Gilbert equation based on the 2D micromagnetic domain wall energy \cite{Thiaville2012}:
	\begin{align}
		\mathcal{E} &= A\left[\left(\frac{\partial\vec{m}}{\partial x}\right)^2 + \left(\frac{\partial\vec{m}}{\partial y}\right)^2\right] + K_u\left[{m_x}^2 + {m_y}^2\right]\nonumber\\
		&\quad -{M_s}\vec{m}\cdot\vec{B}_z -\frac{1}{2}\mu_0{M_s}\vec{m}\cdot\vec{H}_d\nonumber\\
		&\quad  + D\left[m_z\;\text{div}\vec{m} -\left(\vec{m}\cdot\vec{\nabla}\right)m_z\right],
		\label{eq:micromagneticEnergy}
	\end{align}
	
	where $K_u$ is the uniaxial anisotropy constant and ${H}_d$ is the demagnetizing field.
	%To further support the experiments proposed here, micromagnetic model has been employed to compare to the predictions of the Dzyaloshinskii Astroid while also offering some approximate timescale over which the switching process should occur.The 1D micromagnetic simulations were performed in Mumax3 with a cell size of 1x1x1nm.  Material parameters were as follows: $\alpha = 0.02$, $A = 1\times 10^{-11}$J/m, $M_s=800$kA/m, and $K_u = 1.2\times 10^6$J/m$^3$.    Figure~\ref{Fig:Micromag_Explanation}b shows the internal magnetization direction of a DW over time in response to a perpendicular field pulse at three different values of the driving field, $B_z$.  Figure~\ref{Fig:Micromag_Explanation}a shows the magnetic configuration at 50 ns for a range of combinations of DMI and applied field, which was used to identify the points previously shown on the Dzyaloshinskii astroid of Figure~\ref{Fig:Astroid}a. The agreement is remarkable and confirms the analytically proposed mechanism for the case of 1D domain walls.
	
	To determine the steady-state profiles of the 1D domain walls for the relevant range of DMI and $B_z$ values, \eqref{eq:micromagneticEnergy} was evaluated for a 60 $nm$ x 1 $nm$ x 1 $m$ nanostrip with an initialized and relaxed DW wall using a mesh size of 1 $nm$ and a run time of $100$ $ns$. (In this case, the derivatives in y in \eqref{eq:micromagneticEnergy} vanish.) Periodic boundary conditions were enforced along the long edges of the nanostrip, allowing for more realistic calculation of the stray fields. During the calculation, the simulation volume was continually centered on the DW. We used the following material parameters: $\alpha = 0.02$, $A = 1\times 10$ $pJ/m$, $M_s=800$ $kA/m$, and $K_u = 1.2$ $MJ/m^3$. Figure~\ref{Fig:Astroid}b shows the magnetic configuration after $100$ $ns$ for a range of combinations of DMI and applied field, which was used to identify the $B_{SW}$ points overlaid on the Dzyaloshinskii astroid in Figure~\ref{Fig:Astroid}a; simulations involving a finer step size in $B_z$ and a run time of $200$ $ns$---which exceeds the upper limit found for the 1D micromagnetic switching time---were used to more accurately ascertain the switching points. As evidenced in Figure \ref{Fig:Astroid}a, the 1D micromagnetic DW calculations agree remarkably well with the analytically-derived astroid. The systematic vertical shift between the analytical and micromagnetic results is indicative of a higher micromagnetic $K_D$ compared to the analytical value which was used in normalizing $B_z$ for both plots.
	
	Additionally, Figure~\ref{Fig:Astroid}c shows the internal magnetization direction of a DW over time in response to a perpendicular field pulse at three different values of the driving field, $B_z$. In all cases considered in the single chirality regime, the switching process occurs within $100$ $ns$ before reaching a steady state. Repeating the micromagnetic calculations in Figure~\ref{Fig:Astroid}c for the 2D case, we calculated the steady-state profiles of DWs in a 60 nm x 60 nm x 1 nm nanostrip. As indicated by the dashed lines, we observed the predicted chirality behavior for each regime---attributing the finer differences in DW core magnetization between the 1D and 2D cases to the distinct demagnetization fields present in each case. Overall, these results confirm the analytically proposed mechanism for the case of domain walls.
	
	To address the case of a 2D skyrmion, we note that despite the fact a skyrmion's dynamics depend on its radius, $R$, its energy can be well-approximated by analytical expressions that retain the same form that gives rise to the predictions of the astroid. Large skyrmions (where $R > 10t_f$ and $R > 100\lambda_0$) can be successfully described by the DW energy model \cite{Cape1971}, as explained in \cite{Buttner2018}. At the other extreme, the energy of sub--$10$nm skyrmions can be represented by an ``effective anisotropy" model \cite{Bogdanov1994, Buttner2018}, as well as the ``compact skyrmion" model put forth in \cite{Bernand-Mantel2020}. Finally, a more generalized model spanning a wide range of $R$, $D$, and $t_f$ values was developed by \cite{Buttner2018}. In each of these models, the expression for the skyrmion energy has the same dependence on $\phi$ as that of \eqref{eq:sigma} in the $q$--$\phi$ coordinates model:
	\begin{equation}
		E = A\cos{^2\phi} + B\cos\phi + C,\label{eq:rDMISystemEquationClass}
	\end{equation}
	where A and B represent the strength dipolar interactions and DMI, respectively, and all three terms are dependent on $R$. We acknowledge that the magnitudes of $D_c$ and $K_d$ differ between the DW and skyrmion cases; as an example, \cite{Bernand-Mantel2020} predicts that $D_c^{Sk}$ $\approx$ 3$D_c^{DW}$ for small skyrmions. Nevertheless, the skyrmions' energetic landscape remains fundamentally asymmetric as per \eqref{eq:rDMISystemEquationClass}. Moreover, for the skyrmion models that describe large and sub--$10 nm$ skyrmions \cite{Cape1971, Bogdanov1994, Buttner2018, Bernand-Mantel2020}, the radial dependence of $A\cos{^2\phi} + B\cos\phi$ is completely separable; thus, the static equilibrium predictions of \eqref{eq:nG} and \eqref{eq:bG} hold exactly. We therefore expect skyrmions to follow the fundamental predictions of the Dzyaloshinskii astroid, %(which is normalized by $D_c$ and $K_d$) 
	with lability lines separating the achiral, chiral, and precessional regimes as seen in Figure \ref{Fig:Astroid}.
	
	Since the Bloch chirality of Dzyaloshinskii DWs at steady state would be preserved after the driving field is removed, these findings % supports a natural
	pave a promising path towards a non-volatile magnetic memory scheme where information is stored as the Bloch chirality of Dzyaloshinskii skyrmions in FM/HM films. Indeed, such a scheme was first presented in 1974 by Slonczewski where carefully timed field pulses in the precessional regime could switch the chirality of Bloch bubbles hosted by iron--garnet films without any reported DMI \cite{Slonczewski1973}. The key distinguishing feature of our work is that the Bloch chirality of skyrmions could be written using field pulses in the DMI-induced chiral regimes, thereby preventing back-switching due to the difference in restoring torque for the two switching directions.  
	%It is noteworthy that the analysis performed here 
	The above analysis makes a number of reasonable yet noteworthy assumptions about DW switching---in particular, that the process occurs coherently without the formation of vertical or horizontal Bloch lines (VBLs and HBLs, respectively). Li et al.\ found that the presence of VBLs increased with decreasing thickness \cite{Li2019, Li2021}. Moreover, the study in \cite{Slonczewski1973} concluded that HBLs mediated switching in the thick films studied; these results have recently been corroborated \cite{Herranen2017}. These DW substructures would distort the shape of the astroid shown in Figure \ref{Fig:Astroid}a, much like the formation of domains impacts the switching fields given by the Stoner--Wohlfarth astroid.  %Understanding the role of these mechanisms would be greatly supported by micromagnetic models of switching. 
	As such, we encourage investigations the relationship between Bloch lines and Bloch chirality preferences using analytical approaches, micromagnetic simulations, and experimental techniques.
	
	\begin{figure}[!htbp]
		\includegraphics[width  = 0.48\textwidth ]{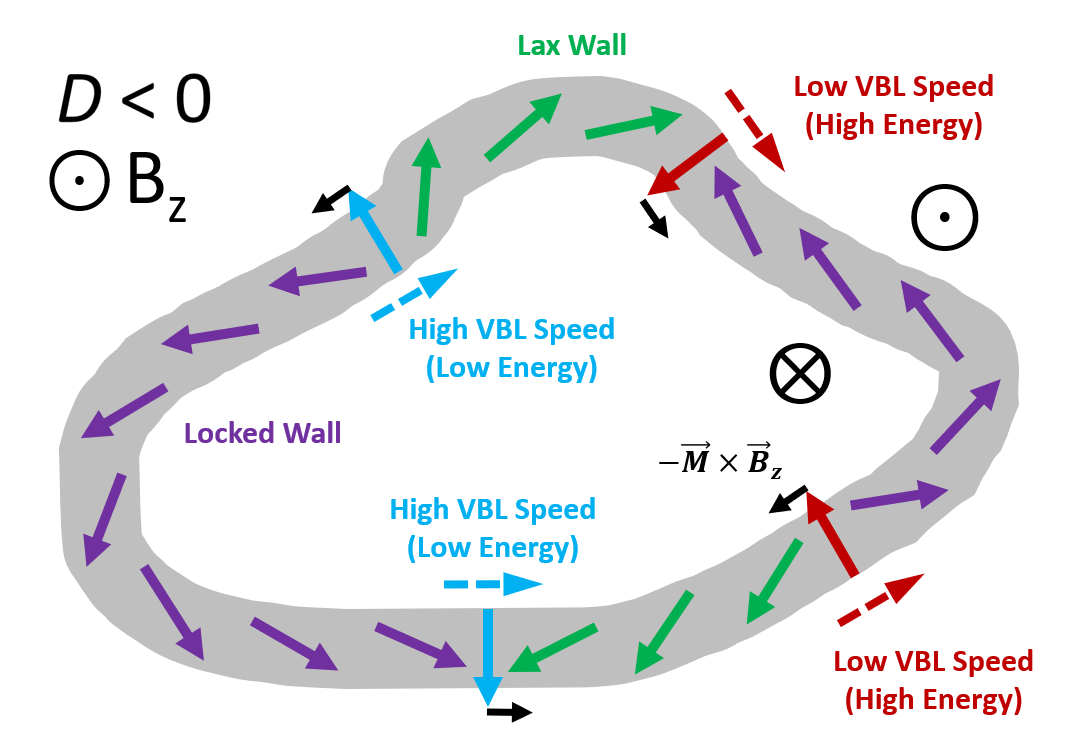}
		\caption{Illustration of an emergent preference for locked walls (purple)---and their Bloch chirality---when VBL nucleation and propagation are jointly assisted by DMI and $B_z$. DMI-stabilized VBLs would nucleate preferentially from lax walls (green) and truncate them. Moreover, DMI-favored VBLs have lower energies and higher field-driven velocities than their counterparts, disproportionately increasing the length of locked walls. \small\label{Fig:AsymmetricVBL}}
	\end{figure}
	
	Interestingly, however, recent research indicates that VBLs can facilitate Bloch chirality preferences in the precessional regime through their interaction with the interfacial DMI field in magnetic thin films. In light of our analysis, the studies presented in \cite{Yoshimura2016, Krizakova2019} suggest that the chiral regime can extend past the Walker field due to DMI-induced asymmetries in the $B_z$-dictated nucleation and propagation of the VBLs that mediate DW precession. As theorized in \cite{Yoshimura2016}, VBLs whose N\'eel components align with the DMI field have lower energies, larger widths and faster propagation speeds than VBLs of the opposite N\'eel handedness. Thus, for sufficiently long DWs, the faster VBLs catch up to the slower ones, ultimately preserving the locked wall configuration and its Bloch chirality.
	
	We also assert that the preference for VBL evolution that is jointly assisted by $B_z$ and DMI would persist at lower fields, permitting VBL-mediated Bloch chirality asymmetries and an extension of the chiral regime below the $B_{SW}$ values indicated by the Dzyaloshinskii astroid. Given an effective $B_z$ and a driving force for VBL formation, DMI-favored VBLs will nucleate preferentially from the lax wall, move faster than the high-energy VBLs, and are more likely to unwind than their counterparts---yielding in each case a preference for the locked wall---and, thus, one Bloch chirality over the other. An example of such a scenario is illustrated in Figure \ref{Fig:AsymmetricVBL}. This behavior is also in line with observed transitions from $\pi$- to 2$\pi$-VBLs in multilayer films with $D < D_c$ as the film thickness decreases \cite{Li2019, Li2021}. Moreover, we anticipate that horizontal Bloch lines would also exhibit similarly asymmetric behavior in the presence of DMI and an out-of-plane field. 
	%it could reasonable lead to a difference in the velocity of vertical Bloch lines that favor one Bloch chirality over the other.
	% effective field needed to induce unidirectional conclusions || Michael: I think we should emphasize that the Bloch chirality was observed in spite of the existence of only interfacial DMI. We would probably also comment on the interlayer DMI theory put forth, simply saying that no microscopic origin was identified to back up its existence or even its purported strength relative to interfacial DMI. || In the context of existing works
	
	Further examining existing literature, we propose that % the mechanism here
	the phenomena theorized above could contribute to the preferred Bloch chiralities % seen in the literature recently. Indeed, a preferred Bloch chirality has been seen in 
	observed in asymmetric Fe/Gd--and Co/Pd--based multilayers with an established or expected interfacial DMI  \cite{Chess2017, Pollard2017, Garlow2019, Pollard2020}. Importantly, these observations were made after various out-of-plane field treatments and in the presence of a non-zero $B_z$. While the authors of \cite{Pollard2020} attributed these findings to an interlayer DMI, the proposed interaction was not probed independent of the observed Bloch chirality distributions, and its microscopic origins were not identified. %Moreover, \cite{Pollard2020} made no direct comparison between the strength of the predicted interlayer DMI and the interfacial DMI which they had determined to be $\approx$ 2.0 mJ/m$^2$ for films with fewer repeats that were otherwise nominally identical to some of those exhibiting a preferentially Bloch chirality.
	%being nominally symmetric (i.e. the z-mirror plane is preserved).  As such, interfacial DMI can only present as a result of differences in the interfaces associated with deposition order---something that has been widely noted for the case of Pt/Co/Pt multi-layers.  
	Superficially, one would not expect gradually applied magnetic fields (as done in these and other studies) to drive the DWs %domains into a steady-state configuration. However, recent work 
	away from static equilibrium as the films quickly adopt stationary magnetic textures.
	Recent work, however, suggests that a millisecond-scale $B_z$ pulse % a slow ramping of a magnetic field can
	in the creep regime can still drive a DW to a steady-state configuration \cite{Brock2021B}%. Consequently, the effective field would be significant during thermally activated movement even though the average effective field is near zero during the overall movement.
	; the effective field, despite its relatively low magnitude, would then be significant during the thermally activated motion of DWs between pinning sites. Likewise, changes in the effective field could temporarily induce steady-state profiles of DWs during the transient expansion or contraction of domains---and for the appropriate $\Delta{B_z}$ values, unidirectionally switching the Bloch chirality of DWs in the process. Finally, an additional study of one film from \cite{Pollard2020} that exhibited a Bloch chirality preference uncovered a prevalence of VBLs, determined that VBLs interact via stray fields to stabilize pairs of DWs with opposite Bloch chiralities, and concluded that further studies of their evolution were crucial to understanding Bloch chirality trends in multilayer films \cite{Garlow2020}. Thus, these results also lend credence to our theory of VBL-mediated Bloch chirality preferences in the low-$B_z$ regime.
	
	In conclusion, we have demonstrated that for DWs driven by out-of-plane fields in interfacial DMI systems, the degeneracy between the two possible Bloch windings is broken at steady state due to DMI-induced asymmetries in the restoring torque experienced by the two  wall variants. Consequently, a new steady-state regime emerges where DWs of only one Bloch chirality are stabilized as determined by the signs of $D$ and $B_z$. Using the Slonczewski equations of motion and the  q--$\phi$ DW energy model, we derived an effective phase diagram for field-driven Dzyaloshinskii DW switching, which we term the Dzyaloshinskii astroid. Moreover, our micromagnetic simulations of 1D and 2D domain walls confirmed our analytical findings. Referring to known analytical and micromagnetic models, we also demonstrated that 2D Dzyaloshinskii skyrmions have the same DMI-induced, fundamentally asymmetric energy landscape as domain walls---and should also follow the predictions Dzyaloshinskii astroid. Recasting recent studies in the context of our work further revealed that the evolution of vertical Bloch lines---and, more generally, DW substructures---in the presence of interfacial DMI and an effective $B_z$ could extend the bounds of the chiral Bloch regime beyond the critical lability fields predicted by our analysis. Beyond shedding light on recent reports of Bloch chirality preferences in interfacial DMI systems, these findings present favorable prospects for new non-volatile computing technologies based on the deterministic, writable Bloch chirality of Dzyaloshinskii DWs and skyrmions.\\

	\begin{acknowledgments}
		
		All authors acknowledge the support of the Defense Advanced Research Agency (DARPA) program on Topological Excitations in Electronics (TEE; grant no. D18AP00011), as well as the use of the Materials Characterization Facility at Carnegie Mellon University supported by the grant MCF-677785. M.D.K. is also grateful for the support of the National GEM Consortium, as well as the support of the Neil and Jo Bushnell Engineering Fellowship awarded by Carnegie Mellon University's College of Engineering.
	\end{acknowledgments}

	\clearpage
	
	%\nocite{*}
	\bibliography{citations}
	
\end{document}